\documentstyle[aps,prbbib,twocolumn,epsf]{revtex}

\begin{document}
\draft
\title{Nonlinear I-V Characteristics of a Mesoscopic Conductor}

\author{Baigeng Wang$^1$, Jian Wang $^1$ and Hong Guo$^2$}
\address{1. Department of Physics, The University of Hong Kong, 
Pokfulam Road, Hong Kong, China\\
2. Center for the Physics of Materials and Department 
of Physics, McGill University, Montreal, PQ, Canada H3A 2T8\\}
\maketitle

\begin{abstract}
We present a general theoretical formulation, based on nonequilibrium 
Green's functions, for nonlinear DC transport in multi-probe mesoscopic
conductors. The theory is gauge invariant and is useful for the predictions
of current-voltage characteristics and the nonequilibrium charge pile-ups 
inside the conductor. We have provided a detailed comparison between the
gauge invariant scattering matrix theory and our theory. We have also 
given several examples where the I-V curve can be obtained analytically.
The effects of exchange and correlation have been considered explicitly.
\end{abstract}

\pacs{73.23.Ad,73.40.Gk,72.10.Bg}

\section{Introduction}

Many practical electronical devices, such as diodes and transistors,
operate on nonlinear current-voltage (I-V) characteristics: 
$I_\alpha=I_\alpha(\{V_\beta\})$, where subscripts $\alpha$ and $\beta$ denote
leads which connect the device to the outside world. Classically one
can predict the I-V curves by solving the coupled equations of classical
electron motion such as the Boltzman equation, and the Poisson equation for 
the electrostatic potential of the conductor, subjecting to the boundary 
conditions that at the asymptotic region of the lead $\beta$, the external
bias voltage is fixed at $V_\beta$. For coherent quantum conductors in the 
mesoscopic regime,  one still must solve the coupled equations but the
electrons are now quantum entities. Clearly the prediction of I-V curves
becomes much more difficult in the quantum situation. As a consequence, 
most theoretical analysis of quantum transport in coherent quantum devices
do not predict I-V curves: they focus on the linear DC conductance
$G_{\alpha\beta}$ which can be calculated from a variety of theoretical 
methods\cite{landauer,datta1}. 
This is then compared with experiments which extract
$G_{\alpha\beta}$ from the measured I-V curves at a vanishing bias voltage:
$I_\alpha=\sum_\beta G_{\alpha\beta}V_\beta$.

Because linear conductance $G_{\alpha\beta}$ does not give the whole picture 
concerning a nonlinear device, it is very important to, theoretically,
go beyond the linear transport regime and predict the whole I-V curve.
At nonlinear situations one must worry about a fundamental physics 
requirement: the gauge invariant condition which dictates that the predicted
electric current should be the same when potential everywhere is shifted by 
a constant amount\cite{but1}. B\"uttiker and co-workers\cite{but1,but5} have
developed a scattering matrix theory (SMT) which satisfies the gauge 
invariant condition and predicted the second order nonlinear conductance. 
When the scattering matrix takes a specially 
simple\cite{but5} form, {\it e.g.} the Breit-Wigner form, the full nonlinear 
I-V characteristics is also obtained\cite{but5}. The key idea\cite{but1,but5},
in order to maintain gauge invariance, is to include the internal potential 
landscape into the analysis which is a result of long range Coulomb
interactions\cite{but2}. Recently this SMT has been extended to predict 
higher order weakly nonlinear conductances\cite{ma1} and connections to 
the framework of response theory has been formalized\cite{ma}. In both 
SMT\cite{but5,ma1} and the response theory\cite{ma}, one calculates the 
nonlinear conductance perturbatively order by order in voltage. Such theories 
make sense for weakly nonlinear situations where the external bias is 
finite but small. Hence practically it is very difficult to compute I-V 
curves for more general situations.

From the nonequilibrium Green's functions
(NEGF)\cite{meir,ng,chen,hershfield,bruder,konig,yeyati}, 
Ref. \onlinecite{stafford}
provided an analysis of I-V curve in the wide-band limit, where
gauge invariance was satisfied by including the internal potential 
phenomenologically through the use of a capacitive charging model. 
The charging model is, however, not fully nonlinear since the 
internal potential is treated linearly in it. It is thus important and 
attractive to further develop the NEGF to the fully 
nonlinear regime for the purpose of predicting I-V characteristics 
of multi-probe coherent quantum conductors. It is the purpose of this work 
to provide such a development: we have formulated a general gauge invariant 
nonlinear DC theory based on NEGF by treating the nonlinear internal potential 
from first principles. Our theory also goes beyond the wideband limit and 
can directly predict the I-V curves and the nonequilibrium charge 
pile-ups\cite{but1,price} 
inside the conductor. The former can be expanded to obtain weakly nonlinear 
conductances which we shall compare with results obtainable from SMT at 
lower orders\cite{but4}; while the latter gives a voltage dependent nonlinear 
capacitance coefficient which is experimentally measurable.
Our theory provides a solid base for further numerical predictions of
I-V curves for complicated coherent quantum device geometries. We will
provide the theoretical formalism in section II. In section III, we will
give detailed comparison between the result of our theory and that of
SMT. We will calculate the I-V curve for a number of examples. The
summary will be given in section IV. 

\section{Gauge invariant formulation}
Let's consider a quantum coherent multi-probe conductor with the 
Hamiltonian 
\begin{eqnarray} 
H &=& \sum_{k\alpha} \epsilon_{k\alpha}(t) c^{\dagger}_{k\alpha} 
c_{k\alpha} + H_{cen}\{d_n,d^{\dagger}_n\} \nonumber \\
&+& \sum_{k\alpha, n} [T_{k\alpha,n} c^{\dagger}_{k\alpha} d_n + c.c.]
\label{Ham}
\end{eqnarray}
where $\epsilon_{k\alpha}(t) = \epsilon^0_{k} + q {V}_{\alpha}$.
The first term of Eq.(\ref{Ham}) describes the probes where 
DC signal is applied far from the conductor; the second term is the general 
Hamiltonian for the scattering region which is a polynomial in 
$\{d^{\dagger}_n,d_n \}$ that commutes with the electron number
operator\cite{stafford} $N=\sum_n d^{\dagger}_n d_n$; the last term gives
the coupling between probes and the scattering region with the coupling 
matrix $T_{k\alpha,n}$. Here $c^{\dagger}_{k\alpha}$ ($c_{k\alpha}$) is 
the creation (annihilation) operator of electrons inside the $\alpha$-probe. 
Similarly $d^{\dagger}_n$ ($d_n$) is the operator for the scattering region. 
As usual\cite{wingreen,datta}, we assume that at time 
$t\rightarrow -\infty$, the system is disconnected. 

The electric current can be written in terms of Green functions
in the usual manner\cite{wingreen,datta} ($\hbar =1$),
\begin{eqnarray}
J_{\alpha}(t) &=& -q \int^{t}_{-\infty} dt_1
Tr\left[G^r(t,t_1) ~\Sigma^<_{\alpha}(t_1,t)\right.\nonumber \\
& & +G^<(t,t_1) ~ \Sigma^a_{\alpha}(t_1,t)
-\Sigma^<_{\alpha}(t,t_1) ~ G^a(t_1,t) \nonumber \\
& & \left. -\Sigma^r_{\alpha}(t,t_1) ~ G^<(t_1,t) \ \right]\ \ .
\label{i1}
\end{eqnarray}
Here the capital $G$'s are various Green's functions (see below) and the
self-energies are defined as follows:
\begin{equation}
\Sigma^{<,r,a}_{\alpha mn}(t_1,t_2) = \sum_k T^*_{k\alpha m} T_{k\alpha n}
g^{<,r,a}_{k\alpha}(t_1,t_2)
\end{equation}

where
\begin{equation}
g^<_{k\alpha}(t_1,t_2) = i f(\epsilon^0_k)
\exp(-i\int^{t_1}_{t_2} \epsilon_{k\alpha} dt)
\end{equation}

and
\begin{equation}
g^{r,a}_{k\alpha}(t_1,t_2) = \mp i \theta(\pm t_1 \mp t_2)
\exp(-i\int^{t_1}_{t_2} \epsilon_{k\alpha} dt) .
\end{equation}

Using Fourier transforms of the various quantities on the right hand side 
of Eq. (\ref{i1}), it is straightforward to obtain,
\begin{eqnarray}
J_{\alpha} &=& 
-iq \int (dE/2\pi) 
Tr \{ \Gamma_{\alpha}(E-qV_{\alpha}) [
(G^r(E,U) \nonumber \\
&-& G^a(E,U)) f(E-qV_{\alpha}) +G^<(E)]\}\ \ .
\label{i2}
\end{eqnarray}
where $G^r(E,U)$ is the retarded Green's function with $U=U({\bf r})$ the 
electro-static potential build-up inside our conductor. In the Hartree 
approximation it is given by\cite{datta1} 
\begin{equation}
G^r(E,U) = \frac{1}{E-H-qU-\Sigma^r}
\label{gr}
\end{equation}
where $\Sigma^r \equiv \sum_{\alpha} \Sigma^r_{\alpha}(E-qV_{\alpha})$
and $\Gamma_\alpha(E)=-2Im[\Sigma^r_\alpha(E)]$ is the linewidth function. 
Within the density functional theory\cite{but7}, we can further include the
exchange and correlation effect, 
\begin{equation}
G^r(E,U) = \frac{1}{E-H-qU-V_{xc}-\Sigma^r}
\label{gr1}
\end{equation}
where $V_{xc}=\delta E_{xc}/\delta \rho$ is the potential due to the
exchange and correlation energy $E_{xc}$ and $\rho$ is the charge density. 
It is worth to emphasize that a most important departure of our theory from the 
previous NEGF analysis\cite{wingreen,datta,stafford} is that we explicitly 
include the {\it internal} potential landscape $U({\bf r})$ into the 
Green's functions self-consistently. This is the crucial step in the 
development of a gauge invariant nonlinear DC theory.
At Hartree level $U({\bf r})$ is determined by the self-consistent
Poisson equation
\begin{equation}
\nabla^2 U = -4\pi \rho=4\pi i q\int (dE/2\pi)G^<(E,U)
\label{poisson}
\end{equation}
Within Hartree approximation, $G^<$ is related to the retarded and advanced 
Green's functions $G^r$ and $G^a$,
\begin{equation}
G^<(E,U) = G^r \sum_{\beta} i \Gamma_{\beta}(E-qV_{\beta})
f(E-qV_{\beta}) G^a\ \ .
\label{lesser}
\end{equation}
Eq. (\ref{poisson}) is, in general, a {\it nonlinear} equation because
$G^{r,a}$ depends on $U({\bf r})$ (see Eq.(\ref{gr})).  Using Eq. 
(\ref{lesser}) we reduce the current into the following form,
\begin{equation}
J_{\alpha} = -q \sum_{\beta} \int (dE/2\pi) Tr [\Gamma_{\alpha} G^r
\Gamma_{\beta}G^a] (f_\alpha-f_{\beta})
\label{final1}
\end{equation}
where we have used the notation
$\Gamma_{\alpha} \equiv \Gamma_{\alpha}(E-qV_{\alpha})$,
$\Gamma=\sum_{\alpha} \Gamma_{\alpha}$,
and 
$f_{\beta} \equiv f(E-qV_{\beta})$.
To make connection with the SMT, we introduce the screened transmission
function

\begin{equation}
A_{\alpha \beta} = Tr[\Gamma_\alpha G^r (\Gamma \delta_{\alpha \beta}
-\Gamma_\beta) G^a]/(4\pi^2)
\label{aa}
\end{equation}
we then arrive at the familiar form of the current in SMT\cite{but5}, 

\begin{equation}
J_\alpha = -2\pi q \sum_\beta \int dE f_\beta A_{\alpha \beta}
\end{equation}

Eqs. (\ref{final1},\ref{gr},\ref{poisson}) completely determines the
nonlinear I-V characteristics of an arbitrary multi-probe conductor, they
form the basic equations of our theory. The self-consistent nature of the 
problem is clear: one must solve the quantum scattering problem (the Green's 
functions) in conjunction with the Poisson equation. It is easy to prove 
that the current expression Eq.(\ref{final1}) is gauge invariant: 
shifting the potential everywhere by a constant $V$, $U\rightarrow U+V$
and $V_{\alpha}\rightarrow V_{\alpha}+V$, $J_\alpha$ from Eq. (\ref{final1})
remains the same. Note in Eq.(\ref{final1}) the quantity $\Gamma$ depends 
on voltage and without such a voltage 
dependence, the gauge invariance can not be satisfied. On a technical side,
Eqs. (\ref{final1},\ref{gr},\ref{poisson}) also form a basis for numerical 
predictions of I-V curves. For instance one can compute the various Green's
functions $G$ and the coupling matrix $\Gamma$ for multi-probe conductors
using tight-binding models\cite{datta1}; and the Poisson equation
can be solved using very powerful numerical techniques\cite{wang2}.

\section{Results}

The main thrust of the previous section (and of this work) is the solution 
of the gauge invariance problem for DC nonlinear transport in general 
terms of bias voltages. In the rest of the paper we shall derive analytical 
expressions for a number of examples which can be solved in closed form. 
At low bias where SMT\cite{but1} and linear response\cite{ma} are
applicable for weakly nonlinear situation, we show that our general formula 
reduces and becomes compatible with them. But for higher bias where
these previous theories are not applicable, our theory becomes an unique
approach for analyzing nonlinear DC quantum transport.  Hence we derive the
general current-voltage characteristics for the entire range of
nonlinearity for a tunneling device, and prove that results obtained with 
or without gauge invariance can differ substantially not only quantitatively 
but also qualitatively.

\subsection{Weakly nonlinear regime}

For weak nonlinearity we can series expand all quantities in terms of the 
small external bias voltage\cite{but1} and obtain results order by order:
this is precisely the approach adapted in SMT\cite{but1} and response
theory\cite{ma}. In this subsection we confirm that our nonlinear theory
indeed reduces to these previous approaches at the weakly nonlinear regime 
where they are applicable. In particular we shall derive analytical
expressions for the local density of states (LDOS) and the second order
weakly nonlinear DC conductance, which are the two interesting quantities 
for weakly nonlinear regime.

In both SMT\cite{but1} and response theory\cite{ma}, LDOS plays a very 
important role. From our NEGF theory LDOS can be easily derived from the
right hand side of Eq. (\ref{poisson}), which is the charge density,
with the help of Eq. (\ref{lesser}). Here we shall present the explicit 
expression at the lowest order\cite{ma} expansion in the external voltage.
Hence we seek the solution of $U({\bf r})$ in the following form,
\begin{equation}
U= U_{eq} + \sum_{\alpha} u_{\alpha} V_{\alpha} +\frac{1}{2}\sum_{\alpha 
\beta} u_{\alpha \beta} V_{\alpha} V_{\beta} + ...
\label{char}
\end{equation}
where $U_{eq}$ is the equilibrium potential and $u_\alpha({\bf r})$, 
$u_{\alpha \beta ..}({\bf r})$ are the characteristic 
potentials\cite{but1,ma,ma1}. It can be shown that the characteristic 
potential satisfy the following sum rules\cite{but1,ma,ma1},

\begin{equation}
\sum_{\alpha} u_{\alpha} =1
\end{equation}
and 
\begin{equation}
\sum_{\gamma\in\beta} u_{\alpha\{\beta\}_l}=0.
\end{equation}
Here the subscript $\{\beta\}_l$ is a short notation of $l$ indices
$\gamma,\delta,\eta,\cdot\cdot\cdot$. Expanding $G^<$ of Eq. (\ref{poisson}) 
in power series of $V_\alpha$, we can derive the equations for all the 
characteristic potentials. In particular the expansions are facilitated
by the Dyson equation to the appropriate order (in the absence of the
exchange and correlation effect):
\begin{eqnarray}
G^r &=& G^r_0 + G^r_0 \left(qU-qU_{eq} \right. \nonumber\\
& & \left. + \sum_{\alpha} \left[\Sigma^r_{\alpha}(E- qV_{\alpha})   
- \Sigma^r_{\alpha}(E)\right]\right) G^r_0 +\cdots 
\end{eqnarray}
with $G^r_0$ the equilibrium retarded Green's function, {\it i.e.}, when
$U=U_{eq}$. At the lowest order, we thus obtain
\begin{eqnarray}
-\nabla^2 u_{\alpha}({\bf r}) 
&=& 2q^2\int dE f (G^r_0 u_{\alpha} G^r_0 \Gamma G^a_0 
+ c.c.)_{rr} \nonumber \\
&-& 2q^2\int dE f (G^r_0 \partial_E \Sigma^r_{\alpha} G^r_0 \Gamma G^a_0 + 
c.c.)_{rr} \nonumber \\
&-& 2q^2\int dE [G^r_0 (\Gamma_{\alpha} \partial_E f +\partial_E 
\Gamma_{\alpha} f) G^a_0]_{rr}
\label{x1}
\end{eqnarray}
where $\Gamma_{\alpha}$ and $f$ no longer 
depend on voltage after the expansion; and {\it c.c.} denotes complex 
conjugate.

The first term on the right hand side of Eq. (\ref{x1}), which depends
on internal potential $u_\alpha$, describes the induced charge density in 
the conductor. It can be simplified using the fact $i\Gamma =(G^r_0)^{-1} - 
(G^a_0)^{-1}$, hence it becomes $4\pi q^2 \sum_{r'} \Pi_{rr'} u_\alpha(r')$
where $\Pi$ is the Lindhard function\cite{levinson,ma} defined as
$\Pi_{rr'} = -i\int (dE/2\pi) f (G^r_{0rr'}G^r_{0r'r}-G^a_{0rr'}G^a_{0r'r})$.
The second and third term of (\ref{x1}) which do not depend on characteristic 
potential correspond to the charge density due 
to external injection. They are the local partial density of states (LPDOS)
$dn_\alpha({\bf r})/dE$ called {\it injectivity} in the language of the 
scattering matrix\cite{but1}, 
\begin{eqnarray}
dn_{\alpha}(x)/dE &=& 
-\int (dE/2\pi) [G^r_0 \Gamma_{\alpha} \partial_E f 
G^a_0] \nonumber \\
&-& \int (dE/2\pi) f [G^r_0 \partial_E \Gamma_{\alpha} G^a_0  
\nonumber \\
&+& G^r_0 \partial_E \Sigma^r_{\alpha} G^r_0 \Gamma G^a_0 
+ G^r_0 \Gamma G^a_0 \partial_E \Sigma^r_{\alpha} G^a_0 ]
\nonumber \\
\label{inj} 
&=& -\int (dE/2\pi) f [G^r_0 (G^r_0 \Gamma_{\alpha} +\Gamma_{\alpha}
G^a_0) G^a_0] \nonumber \\
&+& \int (dE/2\pi) f [G^r_0 (\partial_E \Sigma^r G^r_0 \Gamma_{\alpha} 
+\Gamma_{\alpha} G^a_0 \partial_E \Sigma^a \nonumber \\
&-& \partial_E \Sigma^r_{\alpha} G^r_0 \Gamma
-\Gamma G^a_0 \partial_E \Sigma^a_{\alpha}) G^a_0]\ .
\end{eqnarray}
Comparing this result with that derived by SMT\cite{but2}, the SMT 
result\cite{but2} corresponds to the first term on the right hand side of 
Eq.(\ref{inj}). Hence the local partial density of states obtained from our
general theory is slightly different from that defined in SMT\cite{but2}. 
However it can be proven that the difference, {\it e.g.} the second
integral of Eq. (\ref{inj}), becomes negligible in the limit of large 
scattering volume. The proof follows the approach detailed in our earlier
work Ref. \onlinecite{zheng}.  

We obtain LDOS $dn({\bf r})/dE$ from Eq. (\ref{inj}): 
\begin{eqnarray}
dn({\bf r})/dE &=& \sum_{\alpha} dn_{\alpha}/dE 
\nonumber \\
&=& -\int (dE/2\pi) f [G^r_0 (G^r_0 \Gamma + \Gamma G^a_0) G^a_0]
\nonumber \\
&=& i\int (dE/2\pi) f (G^r_0 G^r_0 - G^a_0 G^a_0).
\end{eqnarray}
This is exactly the same as the LDOS we obtained from the response 
theory\cite{ma}. The agreement is actually not surprising: because we are 
dealing with DC transport where there is 
time reversal symmetry, the result of NEGF should be the same as that 
of the linear response. Our result of LDOS exactly satisfies the general 
relationship\cite{levinson,but1}: $\sum_{\bf r'}\Pi_{{\bf r}, {\bf r}'}=
dn({\bf r})/dE$. 

As a second comparison to the results obtainable from SMT and response
theory, we now derive a formula for the second order nonlinear
conductance $G_{\alpha \beta \gamma}$ from our theory of Section II. In
weakly nonlinear regime, the electric current can be expanded in a
series form in terms of external bias voltages, 

\begin{equation}
J_{\alpha} = \sum_{\beta} G_{\alpha \beta} V_{\beta} + \sum_{\beta
\gamma} G_{\alpha \beta \gamma} V_{\beta} V_{\gamma} +\sum_{\beta
\gamma \delta} G_{\alpha \beta \gamma \delta} V_\beta V_\gamma V_\delta
+ ...
\label{current}
\end{equation}
To compute the second order nonlinear conductance $G_{\alpha \beta \gamma}$, 
it is enough to calculate $u_{\alpha}$ for the internal potential. Expanding 
the general expression for the current Eq.(\ref{final1}) to the second order 
in voltage, it is straightforward to obtain the second order nonlinear 
conductance:

\begin{eqnarray}
G_{\alpha \beta \gamma} &=& -q^3\int (dE/2\pi) Tr\left[ \partial_E
\Gamma_{\alpha} G^r_0 (\Gamma \delta_{\alpha \gamma}
-\Gamma_{\gamma}) G^a_0 \partial_E f \delta_{\alpha \beta} 
\right.  \nonumber \\
&-& \Gamma_{\alpha} G^r_0 (u_{\beta} -\partial_E \Sigma^r_{\beta}) 
G^r_0 (\Gamma \delta_{\alpha \gamma} - \Gamma_{\gamma}) G^a_0 \partial_E f
\nonumber \\
&+& \Gamma_{\alpha} G^r_0 \partial_E \Gamma_{\gamma} G^a_0 \partial_E f
\delta_{\alpha \beta} - \Gamma_{\alpha} G^r_0 \partial_E \Gamma_{\gamma} 
G^a_0 \partial_E f \delta_{\beta \gamma} \nonumber \\
&+& (1/2) \Gamma_{\alpha} G^r_0 (\Gamma \delta_{\alpha \gamma} -
\Gamma_{\gamma}) G^a_0 \partial^2_E f \delta_{\beta \gamma} \nonumber \\
&-& \left. \Gamma_{\alpha} G^r_0 (\Gamma \delta_{\alpha \gamma} -
\Gamma_{\gamma}) G^a_0 (u_{\beta} - \partial_E \Sigma^a_{\beta})
G^a_0 \partial_E f \right]
\label{g111}
\end{eqnarray}
This result is gauge invariant as one can explicitly verify that it
satisfies the gauge invariant condition\cite{but5} $\sum_{\beta} 
(G_{\alpha \beta \gamma} + G_{\alpha \gamma \beta}) = 0$.
This result agrees with that derived from SMT\cite{but5} if we neglect the 
terms involving $\partial_E \Gamma$ and $\partial_E \Sigma$. 
In that case, we obtain

\begin{eqnarray}
&~& G_{\alpha \beta \gamma} = q^3 \int (dE/2\pi) Tr[(
G^a_0 \Gamma_\alpha G^r_0 u_\beta G^r_0+
G^a_0 u_\beta G^a_0 \Gamma_\alpha G^r_0 \nonumber \\
&-& 1/2 G^a_0 \Gamma_\alpha G^r_0 G^r_0 \delta_{\beta \gamma}  
- 1/2 G^a_0 G^a_0 \Gamma_\alpha G^r_0 \delta_{\beta \gamma})
(\Gamma \delta_{\alpha \gamma} - \Gamma_\gamma)] \partial_E f 
\nonumber \\
&=& 2\pi q^2 \int dE (-\partial_E f) [1/2 q\partial_E A_{\alpha \beta}
\delta_{\beta \gamma} +\partial_{V_\beta} A_{\alpha \gamma}]
\label{g111n}
\end{eqnarray}
which agrees exactly with the result in Ref.\onlinecite{but5}. 
The second order
non-linear conductance has been investigated numerically for several
systems using the scattering approach\cite{wang1}. For detailed
discussion of the numerical technique needed for the calculation, see 
ref.\onlinecite{wang1}.

Finally, we comment that by expanding Eq.(\ref{poisson}) to higher
order in terms of voltage, it is straightforward to show that the 
{\it nonlinear} characteristic potentials $u_{\alpha \beta ...}$ satisfy 
a Poisson equation similar to (\ref{poisson}) with different source terms
$dn_{\alpha \beta ...}/dE$ which correspond to {\it nonlinear} LDOS.
The higher order nonlinear coefficient $G_{\alpha \beta ...}$ can be
obtained in similar fashion as we have done here for the second order
coefficient.

\subsection{I-V curve in the wideband limit at zero temperature}

In the last subsection we examined the limit of weak nonlinearity.
However the main advance we have obtained from the gauge-invariant
NEGF formalism developed in Section II is to be able to predict the full
nonlinear current-voltage (I-V) curves.  Several analysis will be presented
in this and the next subsections for the I-V curves.

In the commonly used wideband limit\cite{wingreen}, the coupling matrix 
$\Gamma$ is independent of energy which drastically simplifies the algebra.
The wideband limit corresponds to cases where the probes have no feature, 
thus the internal potential $U({\bf r})$ becomes just a space-independent
constant $U_o$ (the value of $U_o$ depends on the voltages $\{V_\alpha\}$). 
In wideband limit the steady state Green's function takes a very simple form,
$G_0^r=1/(E-E_0+i\Gamma/2)$, thus the integral in Eq. (\ref{final1}) can 
be done exactly. We obtain,
\begin{equation}
J_\alpha = -\frac{q}{\pi\Gamma} \sum_\beta (\Gamma \delta_{\alpha \beta}
- \Gamma_\beta) \Gamma_\alpha \arctan\left[\frac{\Delta E -qU_o+qV_\beta}
{\Gamma/2}\right]
\label{w1}
\end{equation}
where $\Delta E=E_F-E_0$ and the constant $U_o$ is determined by the charge 
conservation condition $\int dE Tr[G^<(E,V)]=\int dE Tr[G^<(E,0)]$, {\it i.e.}
\begin{equation}
\sum_\beta \Gamma_\beta \arctan\left[\frac{\Delta E -qU_o+qV_\beta}
{\Gamma/2}\right] = \Gamma \arctan\left[\frac{\Delta E}{\Gamma/2}\right]
\label{w2}
\end{equation}
To obtain this equation, the quasi-neutrality approximation\cite{but2}
is assumed which
neglects the charge polarization in the system in addition to the use
of total charge neutrality. The gauge invariant condition in Eq.(\ref{w1}) is 
clearly satisfied: raising both $V_\beta$ and $U_o$ by the same amount does 
not alter the current.  Eqs.(\ref{w1}, \ref{w2}) have been obtained 
before\cite{but5} from SMT where a Breit-Wigner form of the scattering 
matrix is assumed. Hence we may conclude that in this sense the wideband 
limit in NEGF approach is
equivalent to the Breit-Wigner approximation in the scattering matrix approach. 
It is however different from that derived in Ref. \onlinecite{stafford} 
where a linear charging model is used for the internal potential build up. 
It is not difficult to confirm that the result of Ref.\onlinecite{stafford} 
is recovered if we solve for $U_o$ in Eq.(\ref{w2}) to the first order in 
voltage $V$, {\it i.e.} we compute the internal potential $U_o$ by neglecting 
the contributions of higher order characteristic potentials $u_{\alpha 
\beta\cdots}$. In this limit we obtain $U_o = \sum_\alpha 
(\Gamma_\alpha/\Gamma) V_\alpha$. Substitute 
this into Eq.(\ref{w1}) we arrive at the result of Ref. \onlinecite{stafford}.
This exercise also allows us to identify the phenomenological parameter $C_i$ 
of Ref.\onlinecite{stafford} to be $\Gamma_i$, and it indicates that the 
linear charging model for the internal potential is not complete for 
the full nonlinear I-V curve predictions.

Next let's derive the full nonlinear I-V curve for a quantum dot with 
two resonant levels. For two resonant levels in a quantum dot the
retarded Green's function\cite{lin} can be derived to have the following
expression:

\begin{equation}
G^r = \frac{1}{[1/(E-E_1-qU)+1/(E-E_2-qU)]^{-1}+i\Gamma/2}
\end{equation}
where $E_1$ and $E_2$ are energies of the two resonant levels. 
From Eq.(\ref{final1}), we obtain the current is 

\begin{eqnarray}
J_\alpha &=& \frac{q}{2\pi\Gamma} \sum_\beta (\Gamma \delta_{\alpha \beta}
- \Gamma_\beta) \Gamma_\alpha \times \nonumber \\
& & Im[\ln(a_\beta^2-b^2)+\frac{i\Gamma}{2b}\ln(\frac{a_\beta+b}{a_\beta-b})]
\label{i2}
\end{eqnarray}
where $a_\beta=E_F+qV_\beta-qU_o-(E_1+E_2)/2+i\Gamma/2$ and
$b^2=(E_1-E_2)^2/4-\Gamma^2/4$. In the quasi-neutrality approximation
we derive that the internal potential $U_o$ is determined by following
equation,

\begin{eqnarray}
&~& \sum_\beta \Gamma_\beta Im[\ln(a_\beta^2-b^2)+\frac{i\Gamma}{2b}
\ln(\frac{a_\beta+b}{a_\beta-b})] \nonumber \\
&~&=\Gamma Im[\ln(a^2-b^2)+\frac{i\Gamma}{2b}
\ln(\frac{a+b}{a-b})]
\label{u2}
\end{eqnarray}
where $a=E_F-(E_1+E_2)/2+i\Gamma/2$. 

Fig.(1) plots the predicted I-V curve Eq.(\ref{i2}) with the parameter
$\Gamma_1=\Gamma_2=0.1$, $E_1=0.3$, $E_2=1.4$, and $E_F=0.11$. 
To compare the I-V curves with and without gauge invariance, in Fig.(1) 
we have plotted three curves. The dot-dashed line represents 
the gauge invariant solution Eq.(\ref{i2}). Since this solution is
gauge invariant, we choose $V_1=V$ and $V_2=0$ in Fig.(1).
Both the other two lines (solid and dotted) are for $U_o=0$ thus no
internal potential is taken into account self-consistently. Two
observations warrant to be discussed. First, the two non-self-consistent
I-V curves, solid line with $V_1=V$ and $V_2=0$ and dotted line with 
$V_1=V/2$ and $V_2=-V/2$, give different I-V curves. This is clearly
wrong because electric current must only depend on the bias voltage
difference which is $V$ for both curves, and not on the choice of the
reference point for potential. This is a direct consequence of the flaw
of a non-self-consistent theory. 
Second, the
qualitative behavior of the current versus voltage curve is different.
Both curves of the non-self-consistent analysis show quantized steps,
which the self-consistent analysis with quasi-neutrality approximation
does not give. This difference in qualitative behavior can be understood
as due to the quasi-neutrality approximation. The electric current for
the incident electrons with energy $E_F$ will increase sharply when the
chemical potential $\mu=E_F+qV$ is close to the first resonant energy level 
$E_1$. When the internal potential build-up is not included in the
non-self-consistent analysis, this current saturates after crossing the first 
level but increases again when $\mu$ is near the second resonant level. 
This is actually a reasonable picture. In the self-consistent solution
within quasi-neutrality approximation, however, the internal energy $U_o$ 
solved from quasi-neutrality condition Eq.(\ref{u2}) increases linearly 
$V= \kappa U_o$ with the voltage with coefficient $\kappa$ close to one. 
Hence as the chemical potential rises, the resonant level also increases with 
approximately the same amount. Thus the second resonance level will not
be reached (within the quasi-neutrality approximation) for the range of
voltages of Fig.(1). 
Finally, In Fig.(2), we have plotted the differential
conductance $dI/dV$ of two non-self-consistent solutions. For the case
$V_1=V$ and $V_2=0$ (solid line), two peaks show up near $V=0.2$ and $V=1.3$. 
Since the Fermi level in the equilibrium is $E_F=0.11$, those two peaks
reflect the resonant behavior when the chemical potential $qV_1+E_F$ is
in line with two resonant levels $E_1=0.3$ and $E_2=1.4$. When
$V_1=V/2$ and $V_2=-V/2$ (dotted line), the chemical potential is again
$qV_1+E_F=qV/2+E_F$, so we found two peaks at $V=0.38$ and $V=2.6$.
However, the spacing between two peaks are different for two different
choice of voltage $V_1$ and $V_2$: it is therefore important to include
the Coulomb interaction so that the theory is gauge invariant.

Our results strongly suggest that as far as the I-V curve prediction is 
concerned, without interaction one violates gauge invariance, but including 
interaction within the quasi-neutrality approximation is still not enough 
as it misses the expected resonance levels in the I-V curve, which are
often observed in experimental situations\cite{tinkham}. Hence it maybe 
necessary to go beyond the quasi-neutrality approximation.

\subsection{Exchange and correlation effect}

So far the electron-electron interaction which gives rise to the internal 
potential build-up has been treated within the Hartree approximation with the
quasi-neutrality condition.  In this subsection we examine the effects of 
exchange and correlation to the nonlinear I-V curves within the wideband 
limit for a resonant tunneling structure. We must also go beyond the 
quasi-neutrality approximation. 

Going beyond quasi-neutrality approximation means that we must consider the
local charge distribution under the condition of overall charge neutrality.
For a double-barrier tunneling structure, let's introduce capacitance 
coefficients $C_1$ and $C_2$ for the left and the right barrier, 
respectively. Then the charge in the quantum well due to Coulomb 
interaction is given by\cite{but8}

\begin{eqnarray}
\Delta Q&=&-i\int (dE/2\pi) [G^<(E,U_o)-G^<_0]  \nonumber \\
&=& C_1 (U_o-V_1) + C_2 (U_o-V_2)
\label{ex}
\end{eqnarray}
where $\Delta Q$ is the total charge in the well, $U_o$ is the overall
shift of the band bottom of the well due to the Coulomb interaction, and
$G^<_0$ is the equilibrium lesser Green's function. For the system with only 
one resonant level, this equation reduces to 
\begin{eqnarray}
& & \sum_\beta \Gamma_\beta \arctan\left[\frac{E_F-H_0+qV_\beta}
{\Gamma/2}\right] - \Gamma
\arctan\left[\frac{E_F-H_0}{\Gamma/2}\right] \nonumber \\
&=& [C_1 (U_o-V_1)+ C_2 (U_o-V_2)] \pi \Gamma/q
\label{w3}
\end{eqnarray}
where the wideband limit is assumed and $H_0 = E_0 +qU_o +qV_{xc}$. 
The current is given by
\begin{equation}
J_\alpha = -\frac{q}{\pi\Gamma} \sum_\beta (\Gamma \delta_{\alpha \beta}
- \Gamma_\beta) \Gamma_\alpha 
\arctan\left[\frac{E_F-H_0+qV_\beta}
{\Gamma/2}\right]
\label{w4}
\end{equation}

To plot the I-V curve determined by Eqs.(\ref{w3}) and (\ref{w4}), we 
use $V_{xc}=-1.5\alpha \Delta Q^{1/3}$ in the Hartree-Fock-Slater 
approximation\cite{but7,srivastava}, where $2/3 \leq \alpha \leq 1$. 
Parametrized by the coupling constants $\Gamma_i$ and the geometrical
capacitance coefficients $C_i$, we thus can calculate the nonlinear I-V
curves from Eqs.(\ref{w3}) and (\ref{w4}). Fig.(3) presents current as a
function of voltage difference $V=V_1-V_2$.  We have compared the cases 
with or without exchange and correlation potential $V_{xc}$ for two different 
set of system parameters: symmetric barrier with $\Gamma_1=\Gamma_2=0.5$, 
$C_1=C_2=5.0$, $\Delta E=-1.0$, $\alpha=0.7$ (dotted line with $V_{xc}$, 
solid line without); and asymmetric barrier with $\Gamma_1=0.1$, 
$\Gamma_2=0.5$, $C_1=1.0$, $C_2=5.0$, $\Delta E=-1.0$, $\alpha=0.7$ 
(dashed line with $V_{xc}$ and dot-dashed line without). 
We observe that the current for the
symmetric barrier is much larger than that of the asymmetric barrier
for $V>2$, but it can be smaller for smaller bias between $1<V<2$, and
becomes larger again at very small $V<1$. Without the internal potential
build-up taken into account, it is well known that symmetrical tunneling
barriers have larger transmission coefficients hence larger current than
those of asymmetrical barriers. The behavior of the I-V curves in
Fig.(3) at the $V\rightarrow 0$ limit is consistent with this picture. However
at larger voltages this expectation may or may not be true, due to the
nonlinear effects and the internal potential build-up. We found that for
both symmetrical and asymmetrical barriers, the exchange and correlation
effects are to increase the electric current. This is seen more clearly
from the differential conductance $dI/dV$ versus voltage in Fig.(4). Since 
the exchange and correlation term $V_{xc}$ is to lower the bottom of the 
conduction band, the peak of $dI/dv$ shifts to the small voltage as a result. 

When exchange and correlation effects are included in the two-level
tunneling system, the Eqs.(\ref{i2}) and (\ref{u2}) are modified in a similar 
way as Eqs.(\ref{w4}) and (\ref{w3}). Fig.(5) shows the I-V curve for 
two-level system for $\Gamma_1=\Gamma_2=0.1$, $C_1=C_2=1.0$, $E_1=0.3$, 
$E_2=1.4$, $E_F=0.11$, and $\alpha=0.7$. The I-V curves with (dotted line) 
and without (solid line) $V_{xc}$ are plotted for comparison, but both
I-V curves now show two steps reflecting the two resonance levels for
tunneling. Hence by going beyond the quasi-neutrality
approximation, the gauge invariant theory developed in Section II
predicts a "quantized" I-V curve which, as mentioned above, is
physically reasonable. Again, when $V_{xc}$ is present, the
current increases. 

In Fig.(6), we show the differential conductance $dI/dV$ for the 
same system parameters as that of Fig.(5). When the exchange and correlation 
potential is included (dotted line), it is surprising to observe that
there are three peaks in $dI/dV$ instead of two (solid line without
$V_{xc}$). In addition the peaks of $dI/dV$ are shifted towards smaller
values of bias as compared to the $dI/dV$ curve for $V_{xc}=0$. The
entire behavior of $dI/dV$ can be understood as the following. 
The internal Coulomb potential $U_o$ tends to move the resonant level up 
(to higher energy) and the exchange and correlation potential $V_{xc}$ 
tends to lower it down, these two effects give compensating factors to
move the resonance levels inside the quantum well. At very small voltage, 
$|V_{xc}|$ increases much faster than the internal potential $U_o$ does 
as the bias is increased (see inset of Fig.(6)), thus the resonant level $E_1$
moves downwards in energy from its ``bare" value $E_1=0.3$: a resonance
peak is expected when it is lowered to the chemical potential. 
As the voltage increases such that the level $E_1$ is lowered to below
the chemical potential, $dI/dV$ decreases from the resonance peak
consistent with the fact of going off resonance. When the voltage increases
further, the resonant levels $E_1$ and $E_2$ start to move upward in
energy due to the effect of $U_o$, and it will again pass the chemical
potential resulting to the resonance peak near $V=0.5$. Finally when the
voltage is around $V=2.0$, the chemical potential is near the
second resonant level and we see a sharp increase of $dI/dV$ and the
third peak shows up as a result. 
We conclude that the quasi-neutrality condition may need to be extended
in predicting I-V curves when the system is in the tunneling regime. Here
we have used a phenomenological but nonlinear capacitance charging model to
include the charge polarization effects, which are seen to play an important
role in predicting quantized I-V curves.  It is further found that the
exchange-correlation potential $V_{xc}$ can be quite important as it 
provides a compensating effect to the Hartree internal potential $U_o$.  

\subsection{I-V curve with Hubbard U term}

To maintain the gauge invariance, we have so far considered the Coulomb 
interaction in the Hartree approximation with or without an
exchange-correlation term. In this subsection, we will
consider the on-site Coulomb interaction in terms of Hubbard U model
with the following Hamiltonian for $H_{cen}$ in Eq.(\ref{Ham}): 

\begin{equation}
\label{eq3}H_{cen}=\sum_\sigma E_0d_\sigma ^{\dagger }d_\sigma
+U_1n_{\uparrow }n_{\downarrow }
\end{equation}
Here we assume that the quantum dot contains one energy level 
$E_0$ with Coulomb repulsion energy $U_1$ which accounts for the
interaction between different spins. In addition to $U_1$ we assume that 
the long range Coulomb potential $U$ between different sites gives an 
overall a constant shift $U_o$ to the bottom of conduction band. 
This is similar in spirit to the energy shift $\Delta$ introduced in
Ref.\onlinecite{wingreen,li}. However, in our case, $U_o$ has to be 
determined self-consistently. 

The current is still determined by Eq.(\ref{i2}). But now the lesser Green's 
function is given by\cite{ng1}

\begin{equation}
G^<_\sigma(E,U) = -[G^r_\sigma -G^a_\sigma] \sum_\beta \Gamma_\beta
f_\beta/\Gamma
\end{equation}
where\cite{pals,ivanov} 
\begin{eqnarray}
G^r_\sigma(E,U) &=& \frac{<n_{\bar{\sigma}}>}{E-E_0-U_1-qU+i\Gamma/2}
\nonumber \\
&+&\frac{1-<n_{\bar{\sigma}}>}{E-E_0-qU+i\Gamma/2}
\label{gr2}
\end{eqnarray}
and
\begin{equation}
<n_\sigma> = -i\int (dE/2\pi) G^<_\sigma
\label{nsigma}
\end{equation}
The internal Coulomb potential $U$ can be determined in terms of the
geometrical capacitances $C_1$ and $C_2$, 

\begin{eqnarray}
& &\int (dE/2\pi) \sum_\sigma G^<_\sigma(E,U)-
\int (dE/2\pi) \sum_\sigma G^<_{\sigma0} = \nonumber \\
& & C_1(U_o-V_1) +C_2(U_o-V_2)
\end{eqnarray}
from Eq.(\ref{gr2}), this condition becomes
\begin{eqnarray}
& &\sum_\beta \Gamma_\beta \sum_\sigma <n_{\bar{\sigma}}>
\arctan\left[\frac{\Delta E-U_1-qU+qV_\beta} {\Gamma/2}\right]  \nonumber \\
&+& \sum_\beta \Gamma_\beta 
\sum_\sigma (1-<n_{\bar{\sigma}}>) \arctan\left[\frac{\Delta E-qU+qV_\beta}
{\Gamma/2}\right] \nonumber \\
&-& \Gamma \sum_\sigma <n_{\bar{\sigma}}> 
\arctan\left[\frac{\Delta E-U_1}{\Gamma/2}\right]
\nonumber \\
&-& \Gamma \sum_\sigma (1-<n_{\bar{\sigma}}>) 
\arctan\left[\frac{\Delta E}{\Gamma/2}\right] \nonumber \\
&=& [C_1 (U_0-V_1)+ C_1 (U_0-V_1)] 2\pi \Gamma/q
\label{U2}
\end{eqnarray}
With the potential $U$ determined this way, the current is finally written as
\begin{eqnarray}
J_\alpha &=& -\frac{q}{\pi\Gamma} \sum_\beta (\Gamma \delta_{\alpha \beta}
- \Gamma_\beta) \Gamma_\alpha \times \nonumber \\
& & (\sum_\sigma <n_{\bar{\sigma}}> 
\arctan\left[\frac{\Delta E-U_1-qU+qV_\beta} {\Gamma/2}\right] \nonumber \\
&+&\sum_\sigma (1-<n_{\bar{\sigma}}>)
\arctan\left[\frac{\Delta E-U_1+qV_\beta} {\Gamma/2}\right])
\label{hj}
\end{eqnarray}
which can be calculated numerically using Eqs.(\ref{nsigma}), (\ref{U2}), 
and (\ref{hj}). In Fig.(7), we have plotted the differential conductance 
versus voltage for $\Gamma_1=\Gamma_2=0.5$, $C_1=C_2=5$, $\Delta E=-1.0$, 
and $U_1=4.0$. As expected, there are two peaks corresponding to two 
different energies $E_0$ and $E_0+U_1$. Since the Coulomb interaction
$U$ increases linearly with voltage $2U \approx V$, the separation 
between the two peaks becomes $2U_1$. 

\section{Summary}

In this work we have developed a general gauge invariant nonlinear DC
transport theory based on the nonequilibrium Green's functions. The main
idea of this development is to self-consistently couple the NEGF with the
proper Poisson equation for the internal potential build-up inside the
mesoscopic conductor. It is the consideration of the internal potential
distribution which has made the NEGF theory gauge invariant.
At various limiting cases our theory predicts results
consistent with those of scattering matrix theory and response theory, 
but our theory allows a general treatment of the full nonlinear DC transport
regime which are, perhaps, impossible for the other formalisms. 
The present theory is natural to allow the inclusion of exchange and 
correlation potential within the density functional formalism.
Hence it is appropriate for transport in multi-probe conductors in the
quantum coherent regime, and we have applied it to the analysis of resonant
tunneling with one and two resonance levels.  Our results clearly
show that without self-consistent analysis the predicted current would
depend on the choice of potential zero, which is wrong.  For the tunneling
device our analysis also indicated the importance of including charge
polarization effect. This effect can be considered using the
phenomenological model involving capacitance coefficients, as done here; or
it can be included through numerical solutions of a charging model as 
carried out in Ref. \onlinecite{wang2}. Finally we found that in
general a larger current is obtained when exchange and correlation 
effects are included into the analysis.

Many further applications of the present formalism can be made. An
important further development is to abandon the wideband limit. In this
work we have used this limit in order to derive analytical formula, but one
can go beyond this limit in numerical calculations.  In the wideband limit,
the coupling matrix is independent of energy. A consequence, as we observe
from the I-V curves, is that the current increases monotonically with 
bias (no peaks in the I-V curve itself). Hence the negative differential 
resistance (NDR) can not be observed. To overcome this limitation, 
Jauho {\it et. al}\cite{wingreen} have introduced a lower energy 
cutoff to allow a finite occupied bandwidth of the contact. 
This modification allowed NDR to appear, but the current at large 
voltage did not agree with experimental results. Therefore to obtain 
NDR quantitative correctly we must go beyond the wideband limit. This 
can be done within our formalism using the numerical method developed 
by McLennan {\it et. al.}\cite{datta2,datta1}.

\bigskip 
{\bf Acknowledgments.}
We gratefully acknowledge support by a RGC grant from the SAR Government 
of Hong Kong under grant number HKU 7112/97P, and a CRCG grant from the 
University of Hong Kong. H. G is supported by NSERC of Canada and FCAR 
of Qu\'ebec. We thank the computer center of HKU for computational
facilities.

\section*{Figure Captions}

\begin{itemize}

\item[{Fig. (1)}] The current versus the voltage for a resonant
tunneling structure with two resonant levels for self-consistent
solution and non-self-consistent solution with two different
voltage gauges. Solid line: non-self-consistent solution for $V_1=V$ and
$V_2=0$; dotted line: non-self-consistent solution for $V_1=V/2$ and
$V_2=-V/2$; dot-dashed line: self-consistent solution. The system
parameters are $\Gamma_1=\Gamma_2=0.1$, $E_1=0.3$, $E_2=1.4$, and
$E_F=0.11$. 

\item[{Fig. (2)}] The corresponding differential conductance for
the non-self-consistent solutions of Fig.(1). 
Solid line: non-self-consistent solution for $V_1=V$ and
$V_2=0$; dotted line: non-self-consistent solution for $V_1=V/2$ and
$V_2=-V/2$.

\item[{Fig. (3)}] The gauge invariant current versus the voltage for a 
resonant tunneling structure with one resonant level. 
Solid line ($V_{xc}$ not included) and dotted line ($V_{xc}$ included):
$\Gamma_1=\Gamma_2=0.5$, $C_1=C_2=5.0$, $\Delta E=-1.0$, $\alpha=0.7$;
dot-dashed line ($V_{xc}$ not included) and dashed line ($V_{xc}$ included):
$\Gamma_1=0.1$, $\Gamma_2=0.5$, $C_1=1.0$, $C_2=5.0$, $\Delta E=-1.0$, 
$\alpha=0.7$.

\item[{Fig. (4)}] The differential conductance of a resonant tunneling
structure with one resonant level. The parameters are the same as that
of Fig.(3).

\item[{Fig. (5)}] The gauge invariant current versus the voltage for a 
resonant tunneling structure with two resonant levels. 
Solid line ($V_{xc}$ not included) and dotted line ($V_{xc}$ included):
$\Gamma_1=\Gamma_2=0.1$, $C_1=C_2=1.0$, $E_1=0.3$, $E_2=1.4$, $E_F=0.11$.

\item[{Fig. (6)}] The differential conductance of a resonant tunneling
structure with two resonant levels. The parameters are the same as that
of Fig.(5). Inset: the exchange and correlation potential $V_{xc}$ versus
voltage of a resonant tunneling structure with two resonant levels.

\item[{Fig. (7)}] The differential conductance of a resonant tunneling
structure with Hubbard $U$ term. 
The parameters are: $\Gamma_1=\Gamma_2=0.5$, $C_1=C_2=5.0$, $\Delta
E=-1.0$, and $U_1=4.0$. 

\end{itemize}
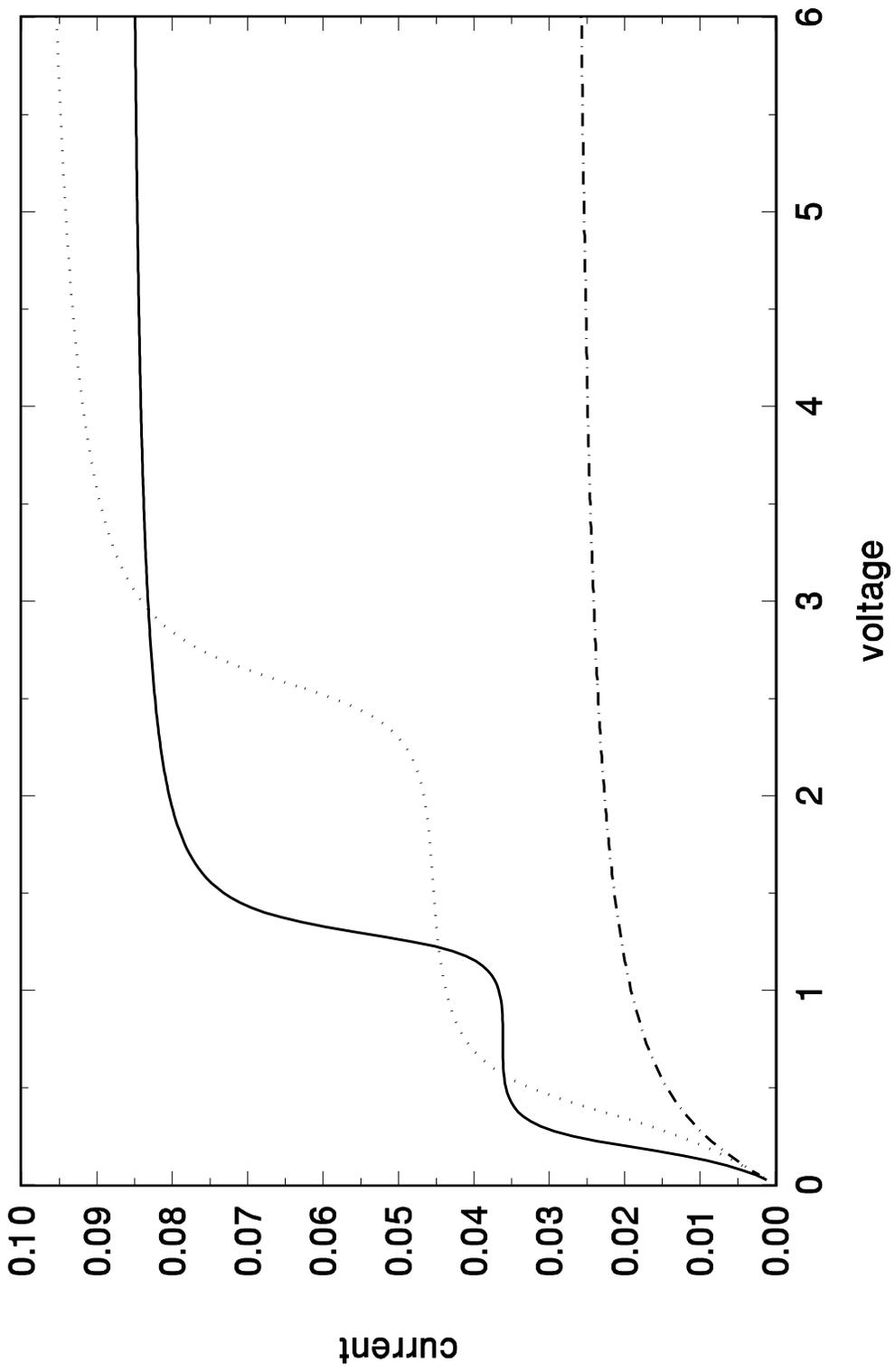
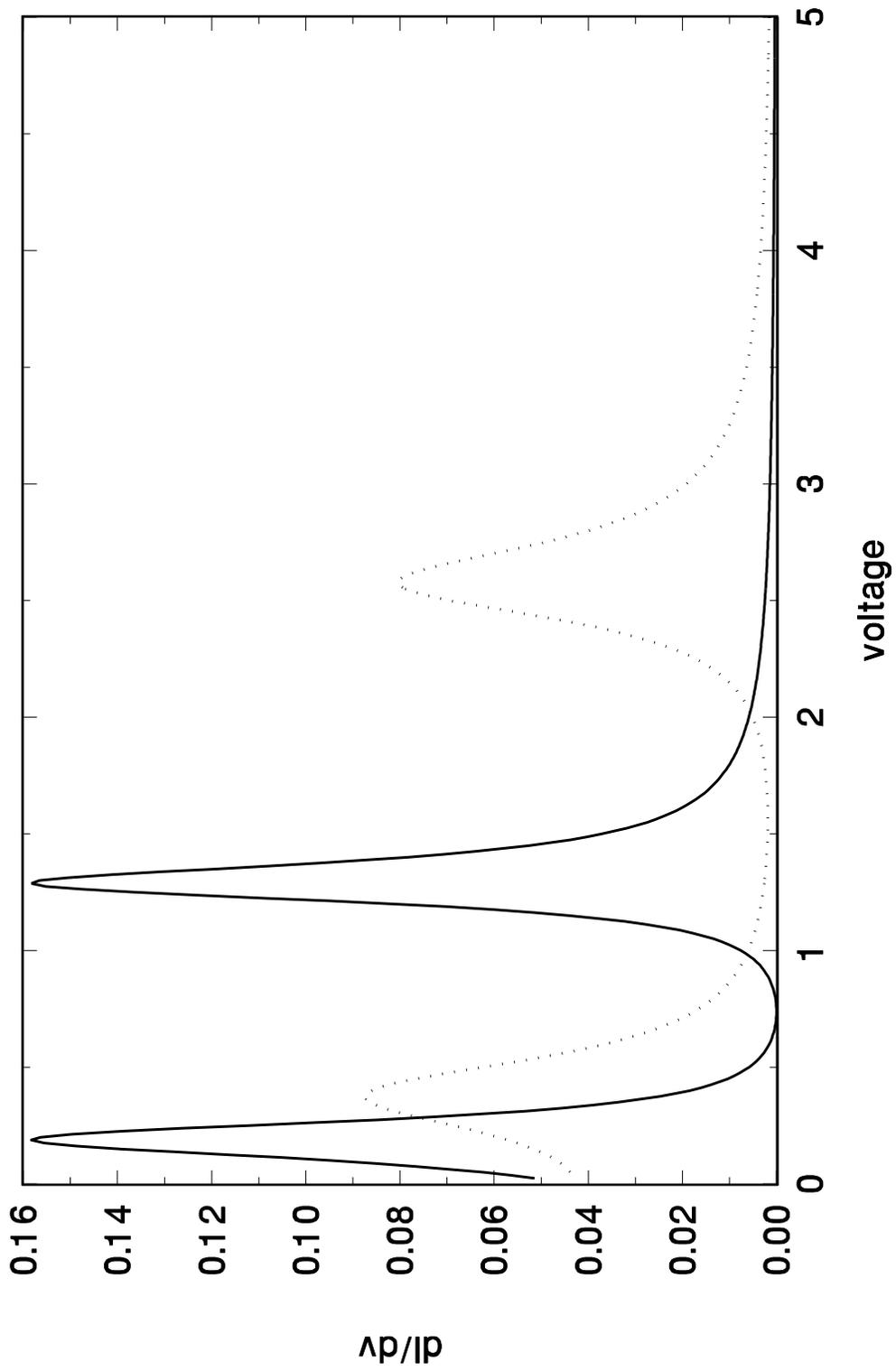
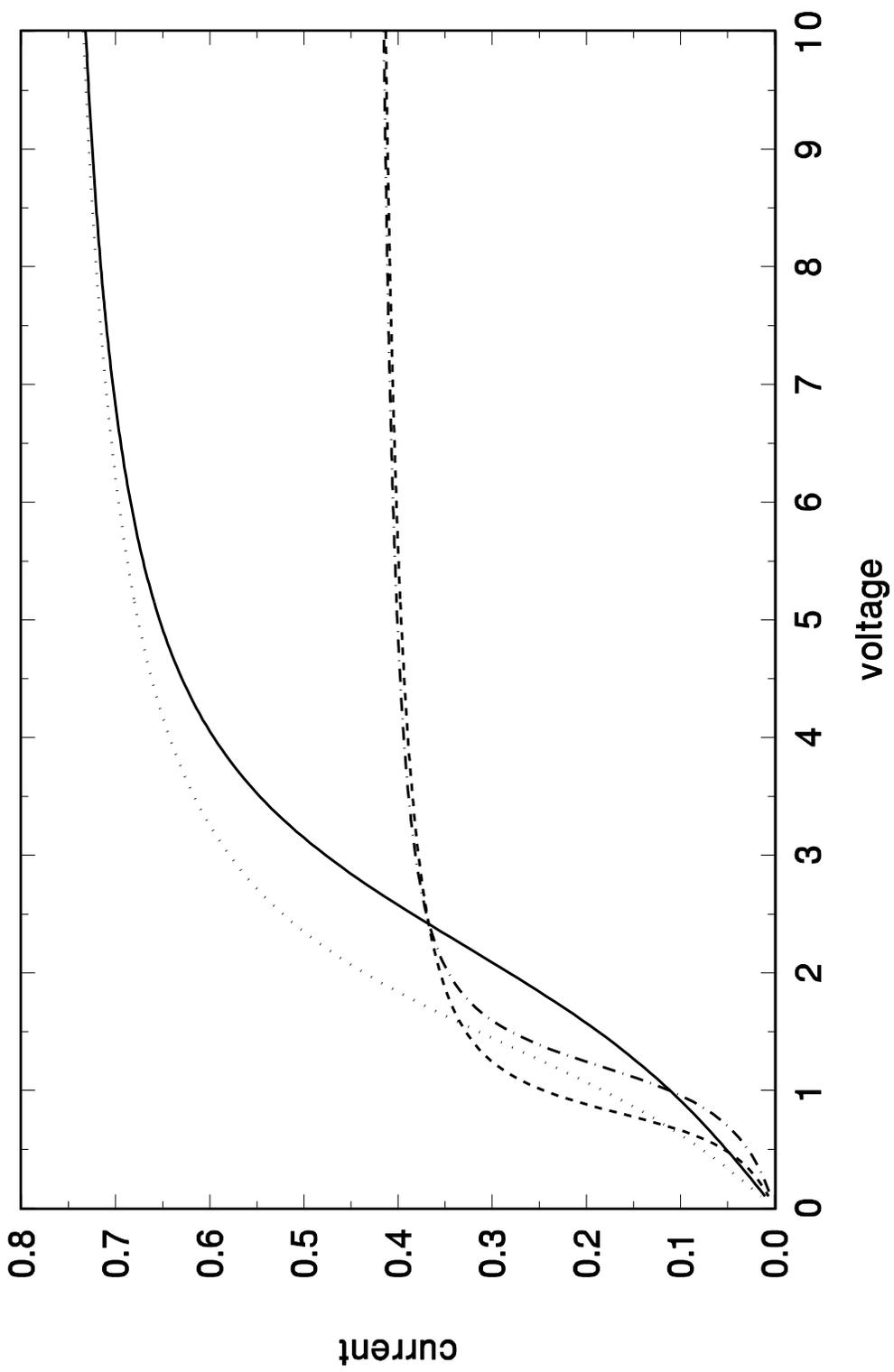
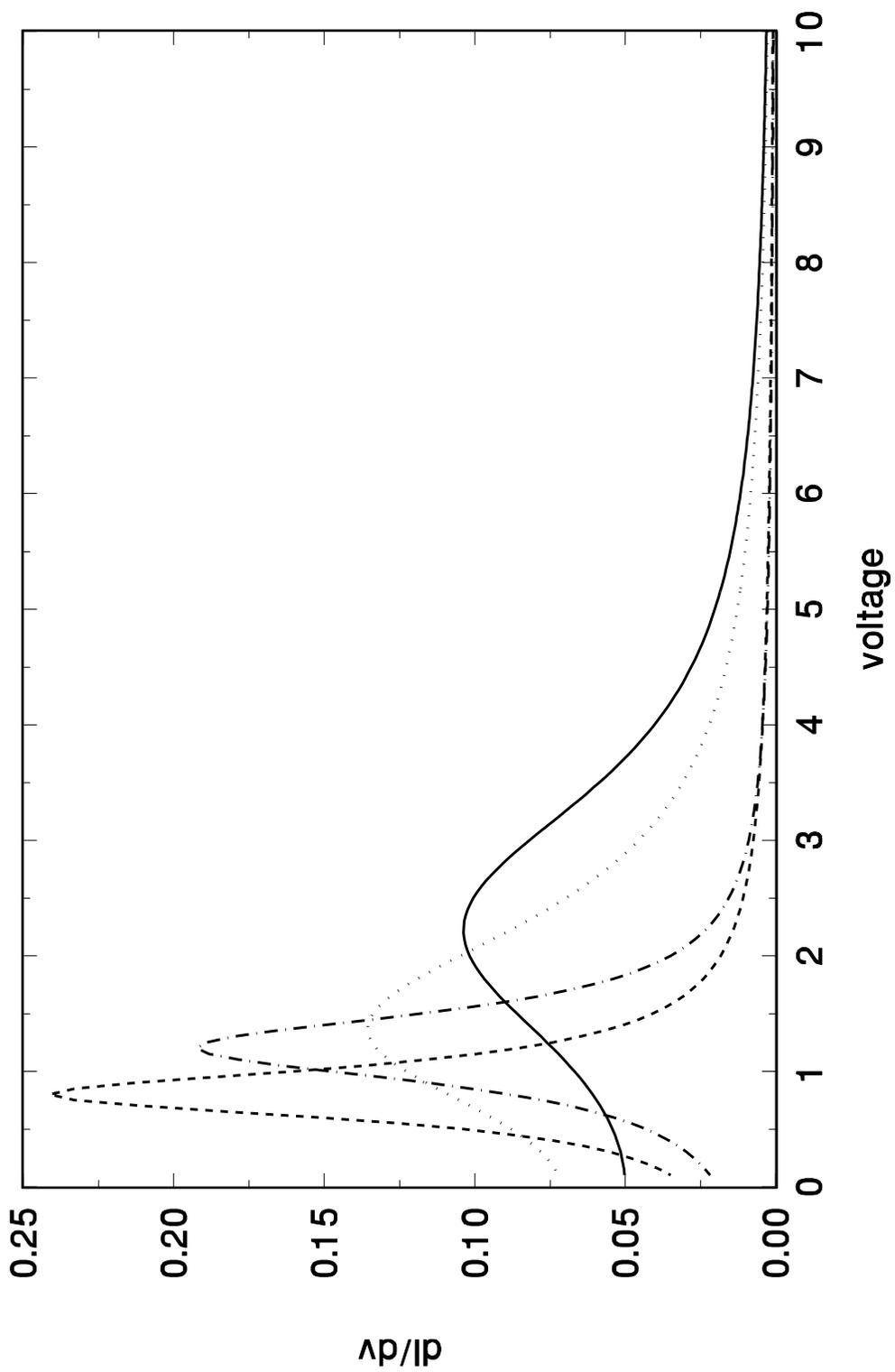
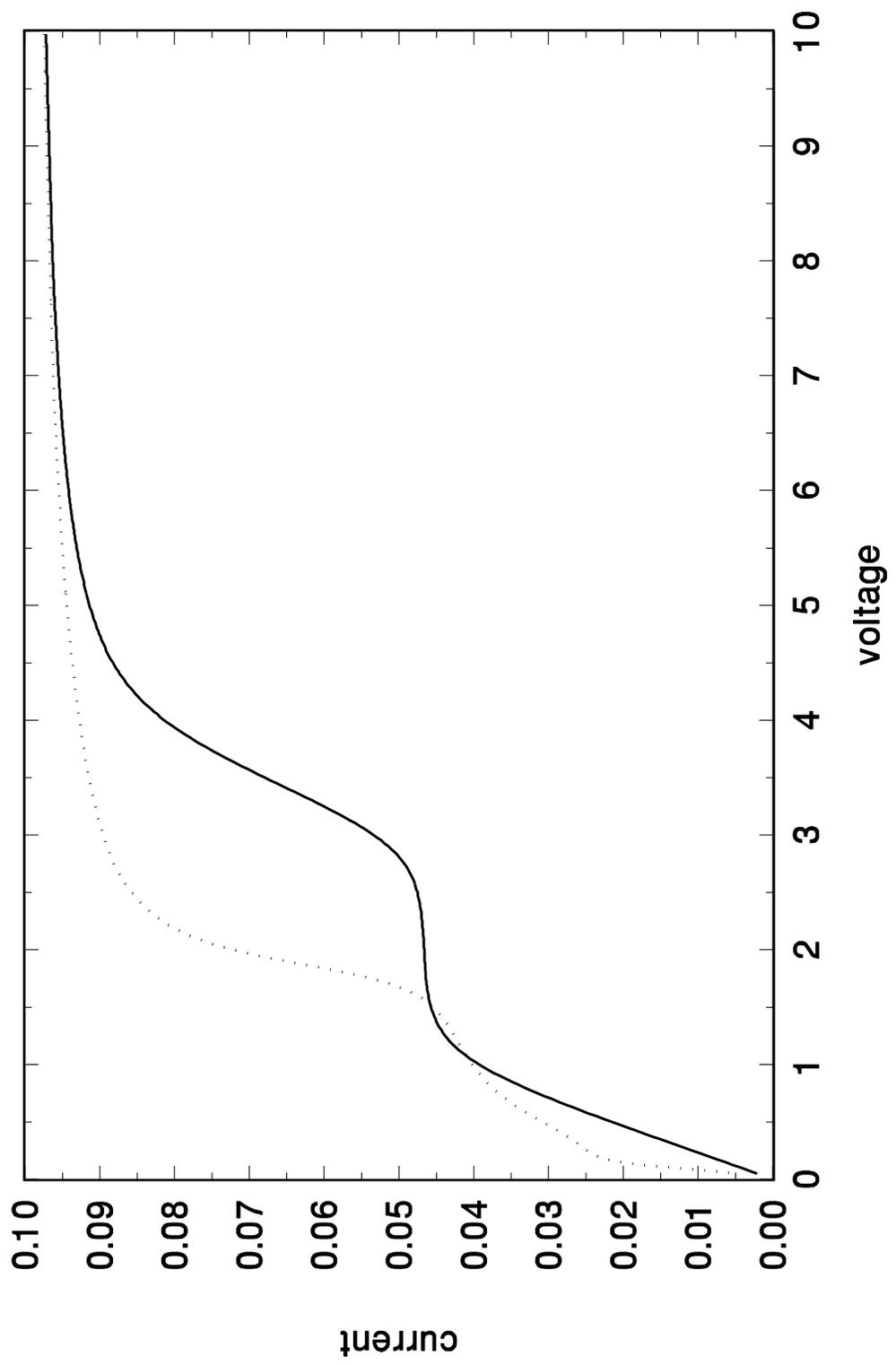
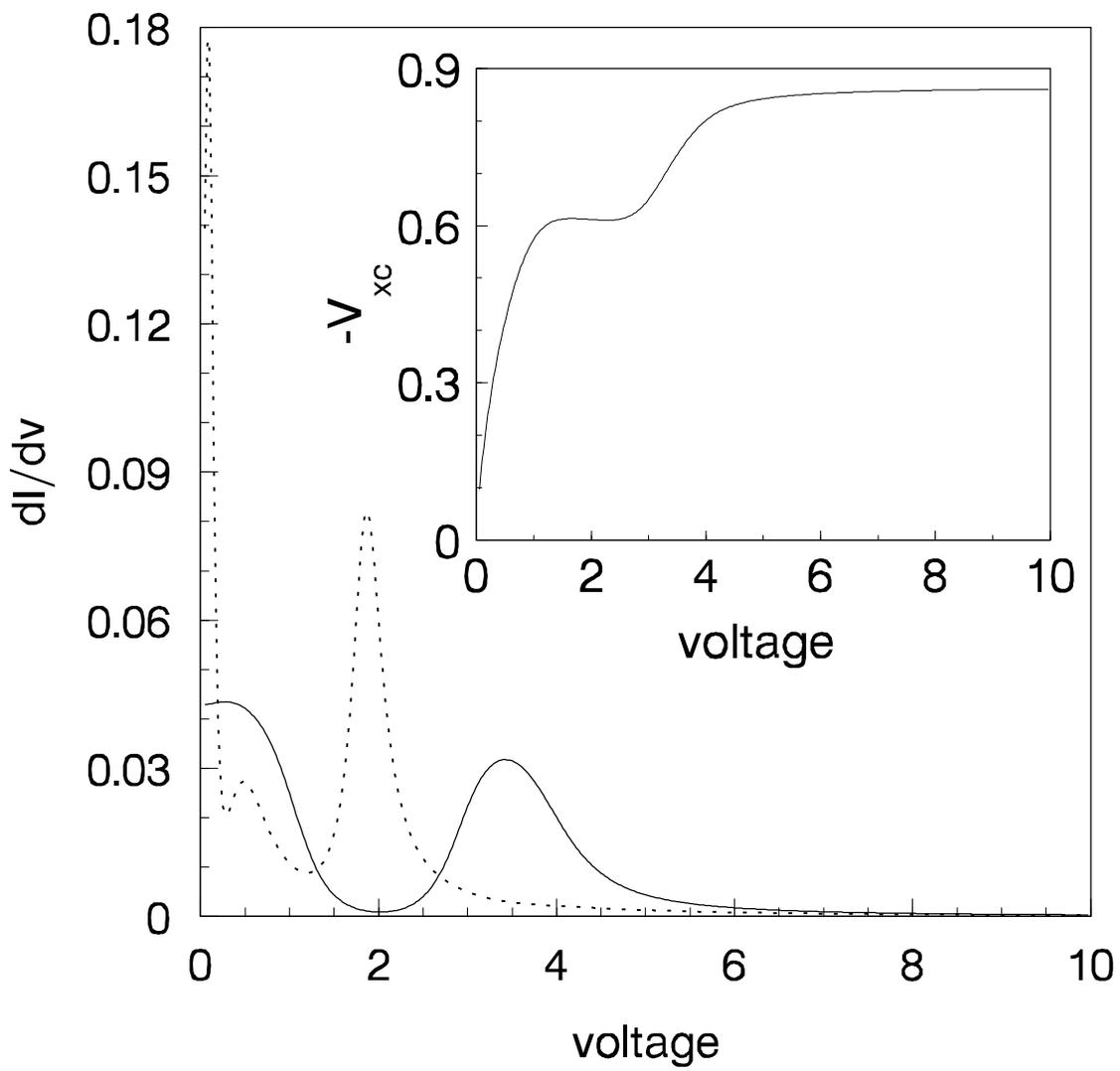
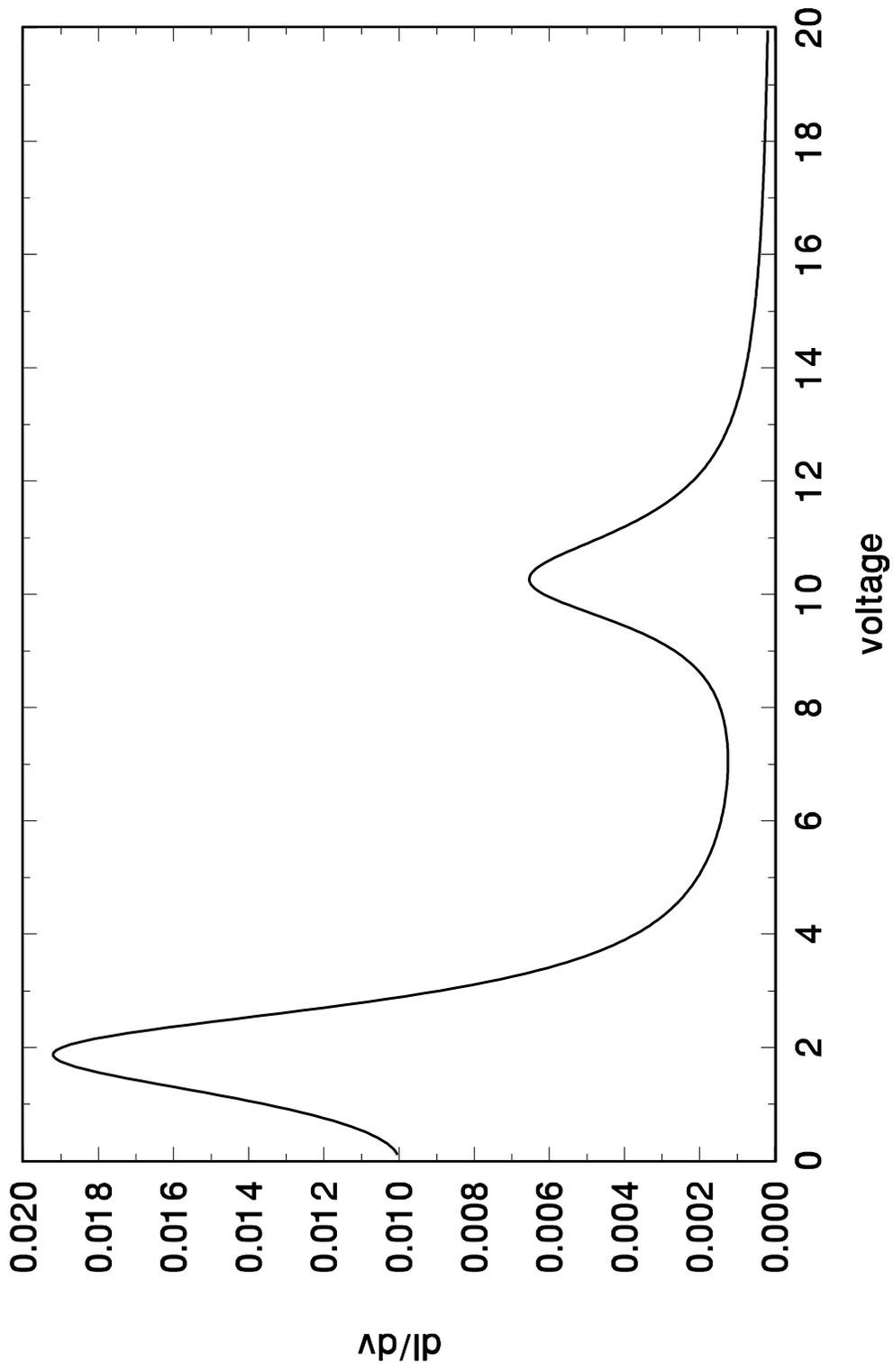
\end{document}